# A review of the sufficient conditions for consciousness.

Peter Coppola, PhD*

## Abstract

How subjective experience (i.e., consciousness) arises out of objective material processes has been called the hard problem. The neuroscience of consciousness has set out to find the sufficient conditions for consciousness and theoretical and empirical endeavours have placed a particular focus on the cortex and subcortex, whilst discounting the cerebellum. However, when looking at neuroimaging research, it becomes clear there is substantial evidence that cerebellar, cortical, and subcortical functions are correlated with consciousness. Neurostimulation evidence suggests that alterations in any part of the brain may provoke alterations in experience, but the most extreme changes are provoked via the subcortex. I then evaluate neuropsychological evidence and find abnormality in any part of the brain may provoke changes in experience; but only damage to the oldest regions seem to completely obliterate experience. Finally, I review congenital and experimental decorticate cases, and find that behavioural evidence of experience is largely compatible with the absence of the cortex. The evidence, taken together, indicates that the body, subcortex and environment are sufficient for behaviours that suggest basic experiences. I then emphasise both the importance of the individual's developmental trajectory and the interdependencies between different neural systems.

*Email: pc605@cam.ac.uk

## 1. The neural correlates of consciousness paradigm

Consciousness is often defined as qualitative subjective experience[1–5]. The concept refers to "what it is like" to be a specific organism in a specific state[6]. This definition is based on a similitude, and is testament of the fact that the objective (extrinsic) evaluation of (intrinsic) subjectivity is problematic. Nonetheless, the evaluation of the presence of experience is an imperative in clinical settings, such as when assessing the depth of anaesthesia[7], or diagnosing disorders of consciousness (e.g., minimally conscious state and unresponsive wakefulness syndrome)[8]. Consciousness is evaluated via observable behaviours of varying degrees of complexity such as movement (automatic, purposeful, functional), verbalisation (intelligent or non; reporting), visual function (startle vs pursuit)[8–10], sometimes called the **behavioural correlates of consciousness**[4]. Although such behaviours may imply the presence of experience, they may not do so infallibly[11]. Similarity, absence of behaviour does not imply absence of consciousness (e.g., locked in syndrome/ cognitive motor dissociation; catatonia; language deficits; cerebellar cognitive affective syndrome etc.,)[7,10,12–14]. In fact, such difficulties are highlighted by the high rate of misdiagnosis of disorders of consciousness[15–17] and intraoperative awareness[18] in anaesthesia.

Partly due to its subjective and non-falsifiable nature (i.e., cannot be unequivocally identified), just more than 30 years ago, it "was heresy to mention consciousness and neuroscience within the same breath"[19]. Yet, now there are numerous yearly publications on the neuroscience of consciousness and a dedicated journal. Such investigations were inaugurated in 1990, when Crick and Koch proposed it was possible to start asking questions in regards to the neural basis of consciousness. They suggested that at any one moment there will be some neuronal processes that will be associated with consciousness, whilst other processes will not be; "What are the differences between them?"[20]. The paradigm underlying the neuroscientific study of consciousness has been further formalised as the search for the **neural correlates of consciousness (NCC)**[4]. The NCC are the "minimum neural mechanisms jointly sufficient for any one specific experience". Specifically, Koch and colleagues propose the research strategy of relating behavioural correlates of consciousness (~responsiveness) to the neural mechanisms underlying them[4]. They further specify the **content-specific NCC**. These are the neuronal processes that determine a specific phenomenological distinction (e.g., a smiling face of a friend). The **full NCC,** on the other hand, refers to the union of all content-specific NCC for all possible contents of experience. This is much more complex, as would include many different contrasting brain states (e.g., dreaming vs awake) and even anatomical configurations (infant vs mature brains, before or after trauma). These authors suggest that contrasting brain activity between conscious and non-conscious individuals (e.g., in unresponsive wakefulness syndrome) and perceptions (e.g., subliminal vs reportable) will yield an identification of the full and content-specific neural correlates of consciousness respectively. Although the concepts of content-specific and full NCCs will be relevant to the review below, unless specified otherwise (as "full NCC" and "content-specific NCC"), the aim of this review is to explore the *minimum requirements for any experience. That is, the neural mechanisms that are sufficient in generating consciousness.*

For the purposes of this review, I simplify the question significantly. I ask **where are the "neural correlates of consciousness"**. Even more simply, I shall ask whether the cortex, subcortex or cerebellum sustain processes that *are sufficient and/or necessary for generating experience*. Of course, although rendering the question more tractable, such a heuristic formulation may be fundamentally ill posed, something which will be discussed at the end of this piece. Nonetheless, the conclusions of this review may provide principled future directions towards a neuroscientific understanding of consciousness.

Many theorists and researchers believe that frontal and/or posterior midline regions of the cortex[3,4,21–24] are important for consciousness. Specifically, modern discussions tend to focus on whether the NCCs are in the "front or the back of the brain"[4,21,22,25–27]. Conversely, the cerebellum is often thought to contribute nothing directly to the experience[3,4,20,28–30]. Furthermore, other theories[31–34] propose phylogenically older subcortical structures to engender experience. Nonetheless, in this review I start with an agnostic stance, and focus on the available empirical evidence.

## 2. Neuroimaging evidence of the NCCs

In neuroimaging research, there are a considerable number of studies focusing exclusively on the cortex[35–47]. These have been successful in showing differences in neuroimaging signals (e.g., electroencephalograph, blood oxygenation level dependent) between unconscious and conscious individuals. Similarly to theory[4,21,25–27], studies implicate both frontal and posterior regions[40,44,48–52] (see figure 1), thus not leading to univocal conclusions as to whether the front or the back of the cortex is essential to consciousness. The cerebellum, conversely, is mostly discounted when it comes to the neuroscience of consciousness[3,4,20,28–30]. However, when investigated, the cerebellum is repeatedly found as a correlate of consciousness across static and dynamic fMRI results[30,50,52–60] (figure 1). In this research, the cerebellum is rarely discussed, but only cited peremptorily as a result, or relegated to the supplementary materials. This evidence suggests that, just like the cortex, the cerebellum's functionality is related to consciousness as it paradigmatically emerges from a contrast between consciousness and unconsciousness.

There are also several studies suggesting that subcortical structures are NCCs[56,61–68]. In particular the thalamus[61–66] and the brainstem[56,61,68] are found to be different in consciousness versus unconsciousness. In diffusion imaging studies, subcortical white matter integrity is able to impressively distinguish between healthy participants and different severities of disorders of consciousness (minimally conscious state and unresponsive wakefulness syndrome)[69] and be predictive of outcome measures[70,71]. Thus, similarly to the cortex and cerebellum, there is evidence indicating subcortical regions are correlated to the emergence of consciousness.

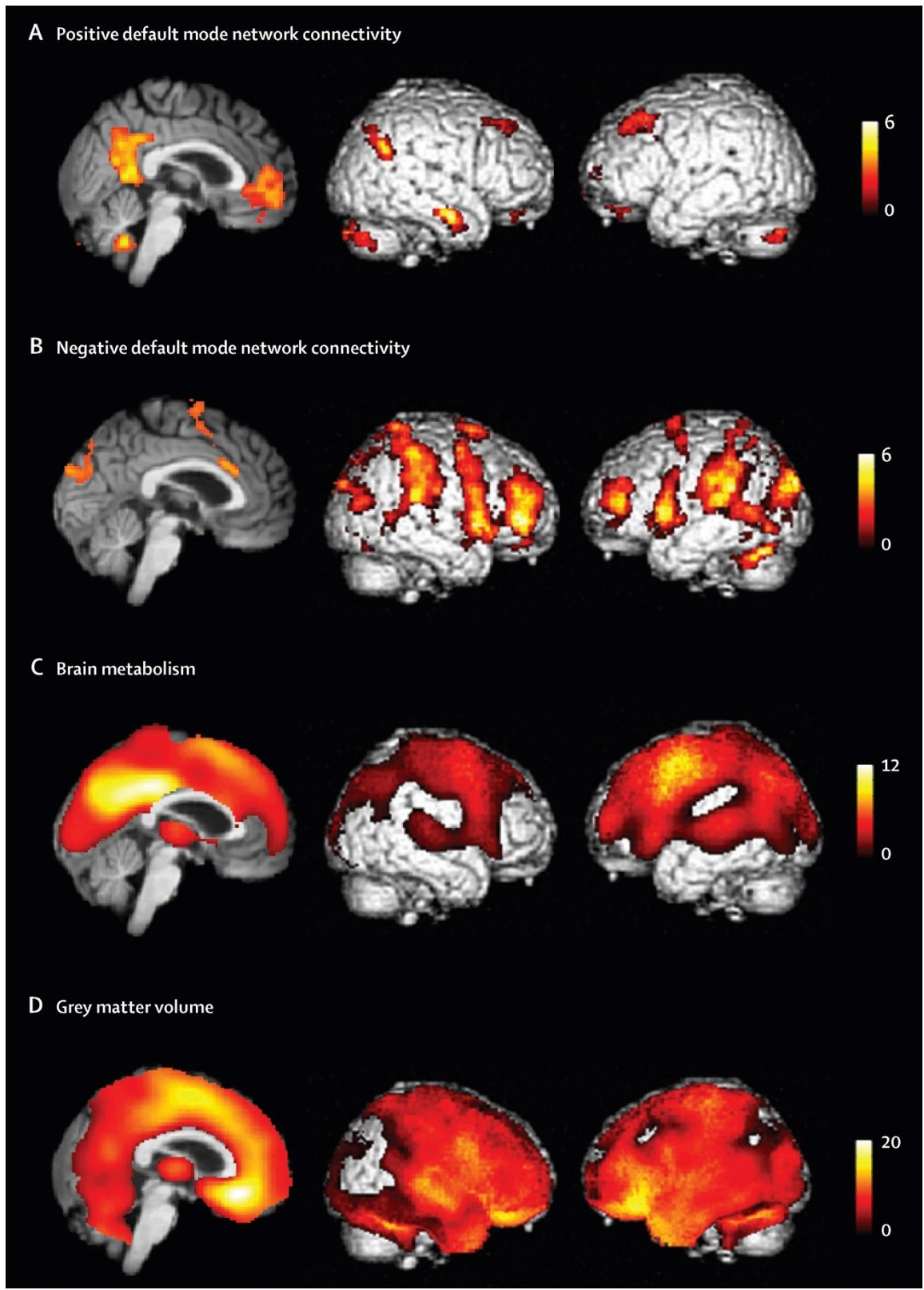

**Figure 1 In vivo brain measures that scale with levels of awareness.** A) positive region of interest (prefrontal and posterior cingulate) functional connectivity (1.4.2). B) negative functional

connectivity from the same regions. Note cerebellar involvement in A and B. C) Brain metabolism as measured by $^{18}$F-fluorodeoxyglucose PET (note posterior midline region's intensity). D) Grey matter volume as measured by voxel-based morphometry (note frontal region's intensity). Colour bars indicate T-values as a function of scaling with presumed levels of awareness (controls>recovering/emergent minimally conscious state> minimally conscious state> unresponsive wakefulness syndrome). Note difference between C and D in posterior and frontal regions. Reproduced with permission from lancet neurology, Di Perri and colleagues[50].

Thus, neuroimaging evidence suggests that the cortex, subcortex and cerebellum are all correlated with consciousness. In view of this, it is possible (1) that the cortex, subcortex and cerebellum are all jointly sufficient for the emergence of consciousness. Alternatively (2), some of these neuroimaging results may be explained as downstream effects of other processes. For example, posterior midline regions are amongst the most complex, interconnected and energy consuming regions of the brain[72]. Hence, the posterior medial cortex may be reliably affected by any interruption of other processes on which it depends (e.g., metabolic, information processing, neuromodulation etc.,). In this case, it would be consistently found as a correlate of consciousness, although potentially being only tangentially related to consciousness itself. Further still (3), it is likely that many neural and cognitive processes are correlated with (or at least modulated by) consciousness. Therefore, the effects observed may be related to functional consequences of consciousness (e.g., decision making, report, prospective thought, mind wondering etc.,) rather than consciousness (i.e., experience) itself[73]. Many of such functional processes are related to the cortex; but are also found to be related to cerebellar function[74–79].

In summary, neuroimaging evidence does not reveal whether cortical, subcortical and cerebellar effects reflect changes in functions associated with consciousness (e.g., cognitive processes)[73], reflect changes in consciousness-imbuing mechanisms, or if such results are relevant to particular aspects of consciousness (e.g., changes in global workspace; basic phenomenology vs complex self-related processing). Thus, a review of more causal types of evidence is warranted.

## 3. Neurostimulation studies

Experimental manipulations of neuronal activity whilst a subject is awake (and able to report changes in subjective experience) may provide insights into the mechanistic processes underlying experience. There are many such reports over the last 80 years with variations in procedures and methodology[80].

In regards to the "back of the brain"[21], alterations of the function of posterior midline cortical regions could induce great changes in the sense of self (e.g., perception of several entities within himself, in the patient's words "the system I know as 'myself' leaves my body")[81]. Such experiences were recently reproduced in other patients[82]. The authors do not speak of alterations to consciousness, but rather to the sense of self. A complex and coherent sense of self is a very prominent aspect of human consciousness ("full NCC"), but nonetheless not necessarily equivalent to subjective experience. In another study[83], stimulating the white matter bundle connecting the thalami to the posterior cingulate cortex caused the surgical patient to stop responding. They consistently reported (upon reacquiring behavioural responsiveness) that they were "in a dream". Such a stimulation unequivocally produced an alteration in consciousness, described by the authors as a disruption of awareness of the external environment. However, subjective reports indicate that subjective experience was preserved (albeit, highly abnormal).

Conversely to the "back of the brain"[21], transcranial magnetic stimulation to the bilateral dorsolateral prefrontal cortex[84] reduced the ability of participants to evaluate their performance on a forced choice task (although this was still above chance). However, TMS did not affect stimulus classification performance, suggesting these individuals could still perceive stimuli. Thus, it seemed that alteration to the dorsal lateral prefrontal cortex affected only a part of awareness ("meta-cognitive"). Nonetheless, stimulation of the dorsal lateral prefrontal cortex in disorders of consciousness reportedly produced some improvement in behavioural correlates of consciousness[8,85–87], although such effects did not equivalate to a cure and tended to work in less severe forms of disorders of consciousness.

Importantly, a review of subjective reports during intracranial stimulation of the cortex[88] (n=1537 stimulations in 67 patients) suggests that subjective changes by stimulation of the prefrontal cortex and the posterior medial cortex are rare. Conversely, it is relatively easy to evoke subjective reports of altered experience in unimodal systems. This, is rather surprising, given the integratory nature of higher-order regions and the theoretical proposals that integration is fundamental to consciousness[3,28]. However, subjective effects from transmodal cortical stimulation, when present, tended to be more complex.

Of interest is that modification of the cerebellum activity via TMS seemed to have a bearing on content-specific NCCs, such as short term memory of visual sequences[90], motion discrimination[91], emotional perception[92]. A meta-analysis on 44 studies suggests medium effect size on perception accuracy[76]. This may suggest that the cerebellum does have a causal role in reportable perception (~content-specific experience) in neurologically typical individuals. Interestingly, similarly to prefrontal stimulation, transcranial direct stimulation of the cerebellum produced improvements in behavioural and neural markers in minimally conscious patients[93].

Compared to cortical and cerebellar stimulation, subcortical electrical stimulation produces rather extreme behavioural and experiential changes[94]. Examples may include the induction of major depression by stimulation of the left substantia nigra in an adult human[95], extreme forms of pleasure arising out of direct stimulation of the nucleus accumbens[96,97] and complex negative emotions through stimulation of the hypothalamus[98]. Whilst, modern reports suggests that stimulation of the amygdala only gave rare (although extreme) emotional responses[99,100], a review of older studies suggest substantial emotional responses do arise[80]. In treatment of disorders of consciousness, stimulation of the thalamus is associated with improved outcomes for minimally conscious patients[101,102]. Three independent studies[43,103,104] found that direct stimulation of the thalamus in anaesthetised unconscious non-human primates induced cortical markers to return to awake-like states (increased firing rate and changes in frequency power). This stimulation also increased behavioural signs of wakefulness (movement, eye-pursuit). The authors do not comment on the extent to which such increases in behaviour may relate to physiological arousal and awareness[10].

There is also substantial evidence that the brainstem has a causal role in the emergence of consciousness. In a classic study, stimulation of the "reticular activating system" (pontine and tegmental areas), produced wakeful-like widespread cortical activation in anesthetised cats[105]. Furthermore, in rats, stimulation of such regions provokes continuous wakefulness[106] or the re-emergence from anaesthesia or traumatic injury-induced symptoms[107]. These rather strong effects of brainstem stimulation (relative to the cortex and cerebellum) are perhaps not surprising, as the brainstem has long been considered "a background condition" for consciousness; that is, *necessary but not sufficient*[4,20]. Of interest is a report that high intensity stimulation of a small fibre bundle between the anterior insula and the claustrum provoked amnesia, and disruption of volitional behaviour in one human patient with resected amygdala (although some motor response was still present)[108]. Analogously, following extended stimulation of the claustrum, there seemed to be behavioural inactivation and reduced awareness of the external environment in five out of eight cats[109]. However, there are also reports of a patient with this region completely destroyed who display all the signs of subjective experience[110]. A possible explanation of this inconsistency is considered in conjunction with other evidence reviewed below (see section 6).

Either way, similarly to neuroimaging research, these neurostimulation studies suggest that function in the whole brain is related to the full NCC. One among several issues with such data, is that any subjective effect elicited may be ascribed to changes of other components of a larger network[80,89]. However, modulation of activity in different parts of the brain seem to change experience in different ways and to different extents (or in some cases even abolish it). This evidence seems to suggest a primary role for older diencephalic structures of the brain, given the powerful effects stimulations in such regions have. To further explore what may be sufficient for experience to arise, I shall investigate cases of acquired and congenital brain abnormalities.

## 4. Neuropsychological studies

### 4.1. Neuropsychological studies of the cortex

Evaluations of changes in consciousness after acquired brain damage may provide strong evidence as to what is necessary for consciousness. Similarly, although of complex interpretation, the presence of experience given congenital abnormalities may provide clues as to what structures are mechanistically fundamental to experience.

There is a report[111] of 11 patients with unilateral ischemic damage to the precuneus exclusively. Of interest here, alterations to "normal" consciousness were reported, including confusion, inability to recognise themselves, loss of sense of agency, impairment in autobiographical memory, body awareness disorders (e.g., alien hand, fading limb) and hemispatial neglect. Thus, damage to this highly connected and complex region may produce severe alterations in consciousness. There also reports of bilateral damage to the parietal lobe. In this syndrome, the patients can be aware of one specific aspect or parts of a scene, but struggle to have a global awareness and display mnemonic deficits[112]. Nonetheless, damage to these regions did not seem to abolish experience, but alter it.

There are, conversely, several studies on frontal lobe damage. A study[113], utilising patients with frontal lobe damage (n=15) and a masking paradigm (disrupting stimulus presentation), found that these individuals had higher thresholds than controls in being able to detect the target stimuli. The patients could still detect stimuli suggesting content specific NCCs were not eliminated. Of recent interest has been an old report of a human that had his bilateral frontal lobes removed[114]. Despite this, the patient could still display emotionally laden complex behaviours that were recognisable to people who knew him. Again, despite reports of severe cognitive impairment after "the complete obliteration" of the frontal lobes, consciousness seemed to be preserved. However, although cited by some neuroscientists of consciousness as evidence that the frontal lobe is not necessary to consciousness[4,21], others[21,26] suggest that the prefrontal cortex was not completely removed in this case. Another more modern patient report with accompanying video (https://www.jneurosci.org/content/37/40/9593/tab-figures-data) shows a clinician investigating the autonomic reflexes of a patient with bilateral dorsal prefrontal damage[26]. There is a noticeable impairment in normal cognition and perhaps extreme disconnectedness from the external environment. The authors state that this patient shows inherently impaired meaningful and wilful behaviour taking this as evidence of impaired subjective experience. However, of importance is that in this short video the patient did follow more than one command, which, would firmly indicate a minimally conscious diagnosis rather than a diagnosis indicating complete unawareness[115]. In fact, such bilateral frontal damage is clinically associated with abulia ("lack of will")[14] which may be a better description of this individual's subjective experience. The syndrome of this patient seems reminiscent of frontal lobotomy treatments[4], which despite altering patients substantially, seemingly did not affect the capacity to "have" experiences. Nonetheless, there is a report of a case with bilateral extensive frontal agenesis in an 8-year-old child (figure 2). Despite near complete absence of frontal lobes and impairment in abstraction and attention, the child showed relatively normal emotional, language and memory. This evidence suggests that the frontal cortex is not necessary for there to be any type of experience (analogously to cerebellar agenesis below). Similarly to the frontal and parietal cortex, destruction of large portions of the temporal lobe (claustrum and insula[116], and hippocampus[117]) did not reportedly seem to abolish experience.

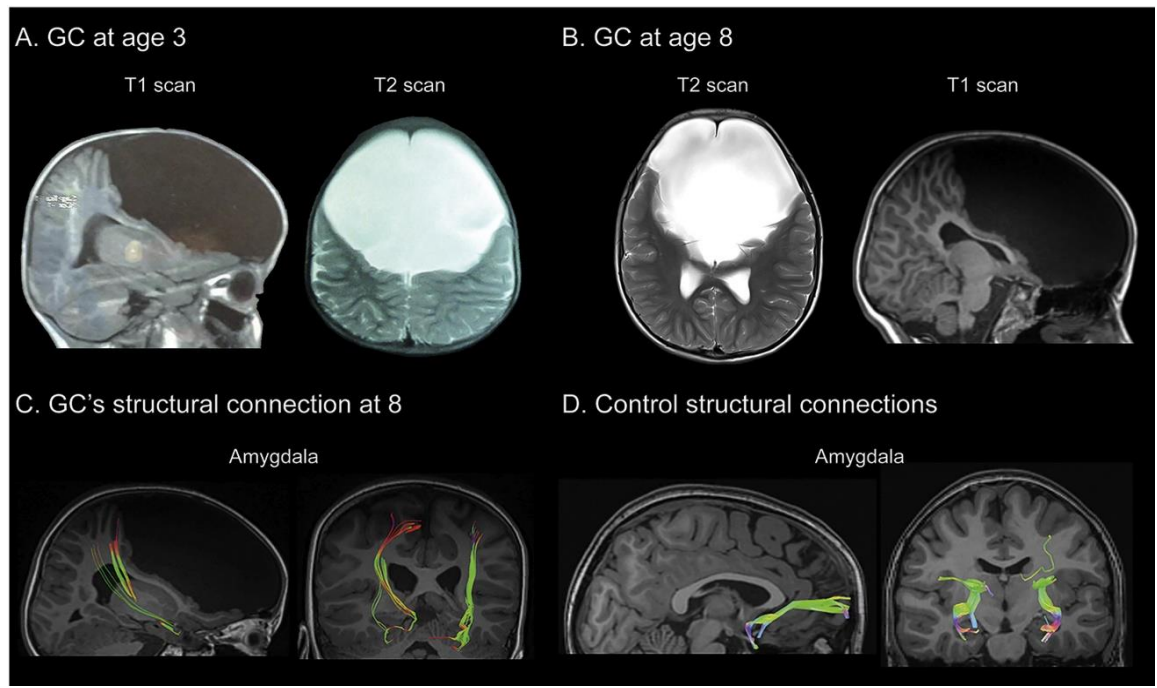

**Figure 2 Frontal agenesis case.** T1 and T2 images (1.3.1) of a child (named GC) with very little frontal cortex (A at age 3; B at age 8). DTI structural connections of the amygdala in GC (C) and in an age and sex matched control (D). Despite very little frontal lobe remaining, this patient showed behavioural vivacity strongly suggestive of subjective experience (see supplementary videos in https://www.sciencedirect.com/science/article/pii/S2213158218300603#ec0005). Image taken lawfully from Ibanez and colleagues, Neuropsychologia[118]

Importantly, authoritative consciousness scientists[4,8] write that the complete destruction of the cortex may induce unconsciousness. They cite from a fundamental clinical text for the clinical practice related with disorders of consciousness[14]. This states that the "diffuse, bilateral destruction" of the cortex is a cause of coma. In support of this, they report a clinical case in which demyelination of the whole brain produced a coma.

However, upon more detailed investigation, the aetiology of reported cases of complete destruction of the cortex seem to be fundamentally metabolic or neurotoxic[14]. Hence, with such fundamental biological requirements not being met, impairment of diencephalic regions may not be excluded. Such reflections are corroborated by another authoritative clinical text on disorders of consciousness[119], which reports, the "surprising" lack of evidence that destruction of the cortex (exclusively) may provoke a disorder of consciousness given the necessary implication of other regions in such metabolic pathologies. There are important considerations to be made on the interdependency between the cortex to the subcortex which are detailed below. Nevertheless, although "diffuse bilateral destruction of the cortex" is classically thought of as a cause of coma (therefore implying that the cortex is necessary to consciousness in a normally developed individual), a careful investigation reveals that the relevant aetiologies are likely affecting regions beyond the cortex.

### 4.2. Neuropsychological studies of the cerebellum.

Although the cerebellum may be found as a correlate of consciousness[30,50,52,54–59], this system is rarely considered in the neuroscience of consciousness. When it is discussed, it is to discredit any role in consciousness it may have[3,4,20,21,28,30]. Nonetheless, there is sufficient neuroimaging evidence[30,50,52,54–59] (reviewed above) that, just like the cortex, the cerebellum's functionality is correlated with consciousness, and as such may be a neural correlate of consciousness.

When consciousness researchers and theorists posit that the cerebellum is not relevant to consciousness[3,4,21] they cite cases of cerebellar agenesis (specifically[120,121]). These are individuals who do not have a cerebellum due to complete congenital failure to develop. Of note is that the effects that cerebellar agenesis has on cognition and normal functioning can be variable and sometimes pronounced (see table 1 in Yu et al.,[121]; also see other cases of cerebellar damage in consensus papers regarding the cerebellum[77–79]). Nonetheless, as far as assessable, these individuals seem to be experiencing, analogously to the frontal lobe agenesis above. Thus, this suggests that indeed the cerebellum is not, strictly speaking, necessary for consciousness.

In cases of acquired rather than congenital cerebellar defects, a study of 8 patients with circumscribed cerebellar lesions revealed impaired emotional face recognition processing, and changes in electrophysiological function[122]. Furthermore, A 38-year-old patient suffered a tumour to the posterior fossa and had rather marked cognitive and affective changes, among other symptoms. After the tumour resection in the cerebellum, there were reports of extreme behavioural changes (diagnosed as posterior fossa syndrome), including hallucinations, uncontrollable burst of emotional behaviour, and disinhibited behaviour[123,124]. The importance of the cerebellum in typical experience is also exemplified by the cerebellar cognitive affective syndrome[14,79,125]. These patients, often displaying isolated focal cerebellar injuries[125], show a wide variety of symptoms, ranging from executive function, working memory, psychosis-like symptoms, and affective changes. Furthermore, in several large consensus papers, integrating findings across many neuroscientific disciplines, there are reports that alterations to the typically developed cerebellum may provoke changes in social cognition[77], verbal working memory[79] and various forms of perception[78].

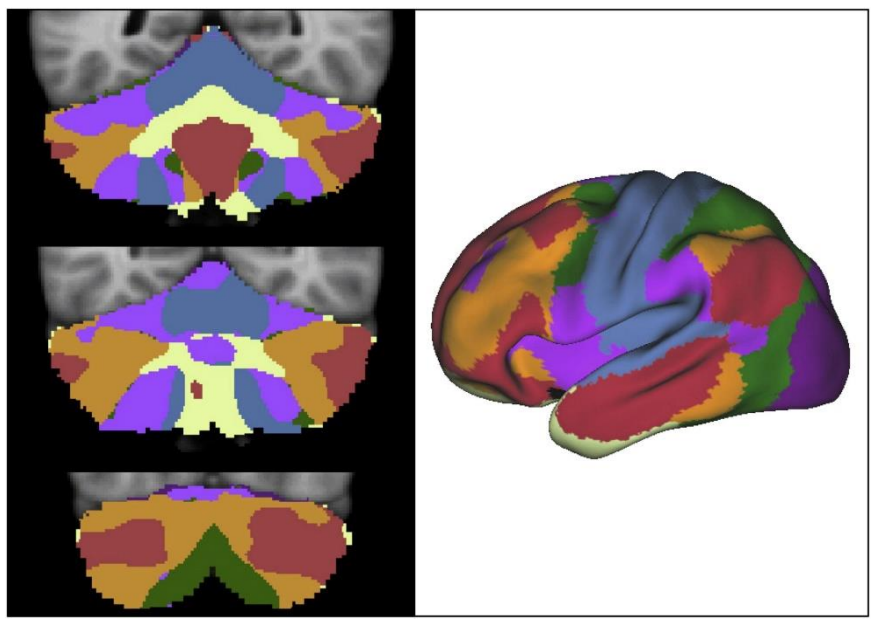

**Figure 3 Functional relationships between cortex and cerebellum.** Functional relationships between cerebellar regions (left) and canonical cortical networks (right) as assessed by functional magnetic resonance imaging. Most of the cerebellum shows functional coupling to higher order cortical regions (Default mode network in red, executive control network in orange). Blue regions indicate somatomotor regions. This may be taken as evidence that the cerebellum has a role in several complex functions. Reproduced with permission from Buckner, 2011.

Thus, despite some cases of cerebellar agenesis with relatively negligible effects (analogously to frontal lobe agenesis[118], figure 2), interruptions of the cerebellum's role in the wider neural system[93,124,126] (figure 3) may potentially provoke substantial changes to behaviour and experience (e.g., affective, perceptual), and therefore may be considered as a part of the full neural correlates of consciousness. Yet, despite a role in specific contents of awareness, and analogously to the cortex (e.g., frontal lobe agenesis ~ cerebellar agenesis), experience was not completely abolished by disruptions of cerebellar functions.

### 4.3. Neuropsychological studies of the subcortex

Subcortical structures are much older than the cortex and are involved in fundamental physiological functions[116,127]. Following on from subcortical stimulation and neuroimaging studies, there is further evidence suggesting that subcortical structures have a fundamental role in consciousness.

The brainstem is routinely implicated in disorders of consciousness[8,14,32,68,116,128,129]. In this clinical context, the destruction of diencephalic (e.g., hypothalamus or thalamus) or brainstem structures are recognised to be causes for coma[14,119]. In fact, brainstem death and brain death are nearly identical when it comes to observable diagnostic criteria[14]. In a study of lesions detectable in MRI, the brainstem (pontine tegmentum, and midbrain peduncles), as well as the caudate, predicted not only unconsciousness, but also recovery from unconsciousness[130]. Parvizi and Damasio[128] showed that, in absence of any other detectable lesions beyond the brainstem, bilateral lesions to the pons or upper pons and midbrain was related with loss of consciousness. Although the lesion mapping overlap between these studies is not perfect, a more modern study[131] replicates these findings. Snider and colleagues, on the other, hand showed that cortical lesions predicted loss of consciousness in as much as they had connectivity with dorsal brainstem regions[132]. Recently[133], computational modelling of biomechanical forces suffered by athletes show that loss of consciousness is particularly associated with higher strain in brainstem regions. Another study[134] found that ischemic-induced, brainstem and thalamic lesions were associated with a severe alteration of consciousness, whist thalamic lesions alone did not have effects of comparable severity. Furthermore, experimentally-induced damage to brainstem regions in animal models is known to consistently provoke sleepiness or coma[14,119,129,135,136], and markedly change cortical activity (resembling sleep or sedation)[137]. A study found that extensive thalamic regions did not produce unconsciousness, but complete bilateral destruction of the basal forebrain, parabrachial and precoreleus brainstem regions was sufficient to produce a comatose state[138]. Cairins reviewed several human neurological cases in which damage to the brainstem or thalamus were likely to precipitate a disorder or consciousness[129]. Bilateral destruction of the hypothalamus in macaques noticeably reduced or abolished emotional behavioural and induced persistent somnolence and drowsiness[139] (similarly to a modern study on mice[140]). Conversely cats with the cortex removed but the hypothalamus intact can display the complete constellation of behaviours that suggest "anger"[141,142].

Penfield and Jaspers[143], were in an exceptionally privileged position in that they diligently probed subjective reports after in-vivo stimulation and resection of hundreds of human brains. They state that "vertical" fibres (i.e., between subcortex and cortex, in humans) are more important to consciousness than horizontal transcortical association fibres. In their clinically driven "centrencephalic" hypothesis, the brainstem is viewed as the super-ordinate system where all information is "re-represented" and ultimately integrated (e.g., to output one motor response despite multifarious information processing in the whole brain). They believed that such loci were "indispensable" to consciousness[136,143], as their disruption would have marked effects on consciousness. Similarly to the modern studies reviewed above (section 3), they were struck by the fact that extensive stimulation and ablation of the cortex "interfered little with consciousness".

## 4.4. Human decorticate cases and consciousness

The evidence reviewed above suggests that, whilst stimulation of the cortex and cerebellum provoke changes in experience, the greatest effects are obtained by stimulating the subcortex. Similarly neuropsychological studies seem to show substantially stronger effects when subcortical areas are damaged. This may be taken to suggest that subcortical structures play an important role in consciousness. However, such a relevance to consciousness may be explainable by the fact (1) that perturbations on the subcortex have an effect on the cortex[43,101,105,143,144] which is required for experience. An alternative explanation (2) is that subcortical structures may be sufficient for the emergence of simple forms of experience[32,94]. Thus, I review "decorticate" cases (with particular attention as to whether such cases display behavioural evidence of having experiences) to attempt to resolve these divergent hypotheses.

Of primary interest are cases of Hydranencephaly. This is a congenital malformation disorder in which most of the cortex is missing (in most cases there are remnants of occipital or frontal lobes)[145,146]. The typical prognosis is irreversible vegetative state (given diffuse bilateral destruction of the cortex, section 4.1) and survival rates can exceed expectations[136,147,148]. In fact, a 32-year-old hydranencephalic woman with little frontal cortex preserved[149] was diagnosed with persistent vegetative state despite groaning vocalisation, demonstrations of facial expressions of joy at the presentation of music, and preserved brainstem auditory function. Despite comparing the patient (by presentation analogy) to a minimally conscious state patient, the author states that "in absence of cerebral hemispheres there can be no consciousness"[149]. Shewmon and colleagues in 1999[147] sustain that such prognoses will engender a self-fulfilling prophecy, as hydranencephalic children's severely affected development would be further stunted by the absence of an adequate environment. Shewmon and colleagues present four hydranencephalic cases (all having variably preserved small remnants of cortex, see figure 4) which were given a relatively normal developmental context by caregivers who had bonded with these children. The authors report behaviours such as the distinguishing of and playing with toys, the enjoyment of and preference for different types of music, distinguishable social responses, anxiety to unfamiliar persons, visual discrimination, a sense of object permanence, and affective responses as well as a basic form of empathy. Despite the presence of minimal cortical matter, in all of these cases, the attending neurologists prognosed with absolute certainty permanent vegetative state, in accordance with guidelines[8,14,115,119]. Similar reports are given elsewhere[136], including reports of different types of learning. In fact, a survey of over 100 hydranencephalic caregivers gave consistent evidence of certain types of awareness (e.g., recognizing family members 91% positive), but, importantly, not others (e.g., behavioural localisation to bodily pain 91% negative)[148].

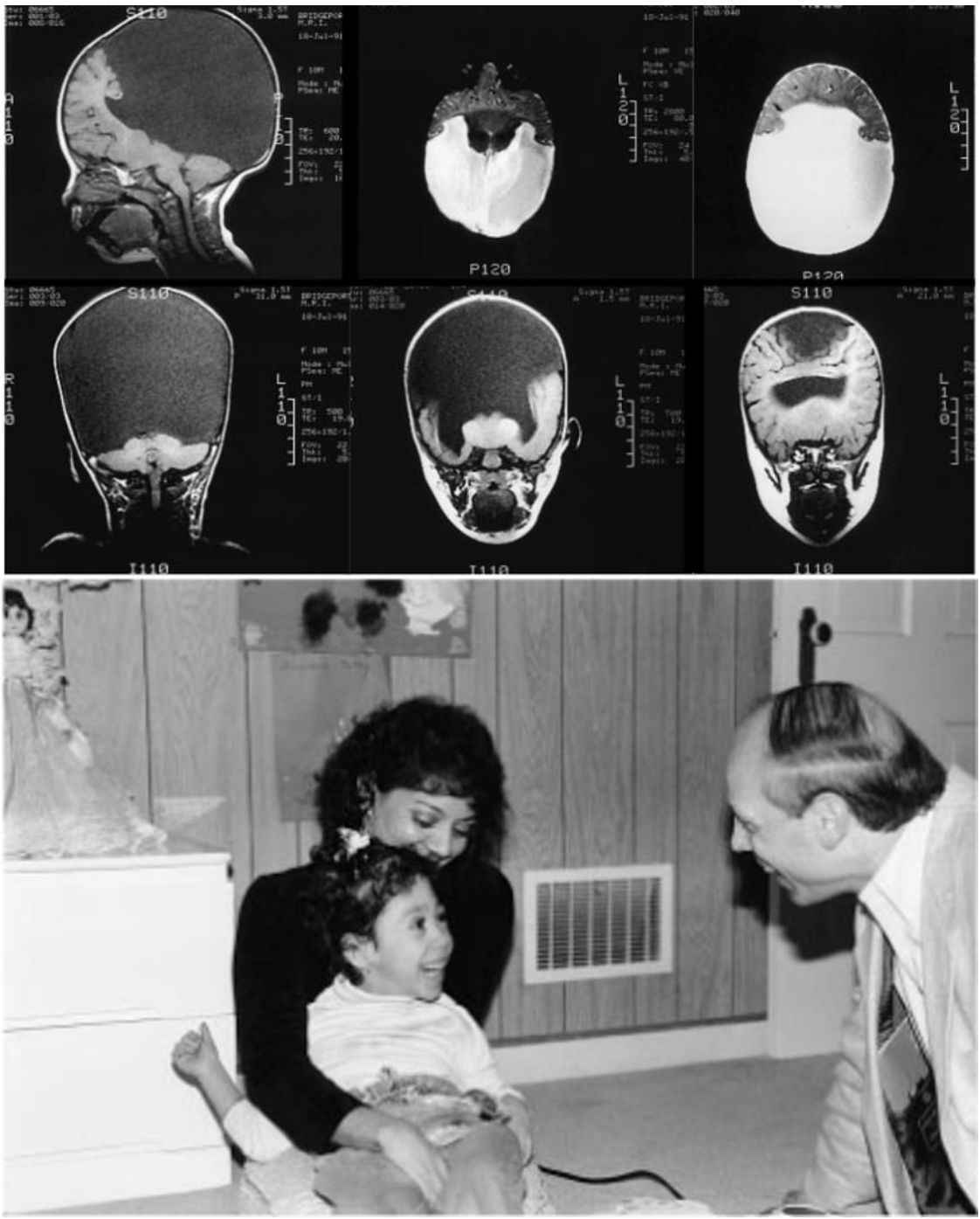

**Figure 4. Subject 3 from Shewmon et al., 1999.** In the upper panel, from left to right, top to bottom, midline sagittal (t1 images), lower and higher axial (t2 images), and coronal planes from anterior to posterior (t1). In the lower panel subject three interacting with an author of the article[147]. Images obtained with permission.

Interestingly, caregivers can recognise the onset and ending of *absence* epilepsy in hydranencephaly, which implies the presence of experience[136,147]. As Shewmon and colleagues point out[147], any ethnologist would not have difficulty to ascribe basic consciousness (e.g., experience of pain) to the behaviour of such organisms. The rarity of such cases is partially due to the fact that the caregivers usually do not disregard the prognosis of a neurologist (i.e., permanent unresponsive wakefulness syndrome/vegetative state)[14,115]. Nonetheless, in most cases, such individuals tend to have remnants of cerebral cortex. The argument could be made that experience is mechanistically mediated by what is left of cortical structures. However, preserved cortical tissue is highly abnormal[146] and potentially non-functional[136]. To my knowledge, there is one modern hydranencephaly case that is reportedly absolutely decorticate[145] (as per neuroradiological assessment, see figure 5). This three-year-old female presented with "severe neurodevelopmental delay", flattened and slow EEG, an observable difference between wakefulness and sleep and was unable to walk without support. Clinicians who observed the child state she was not in a vegetative state and "definitely felt emotions" (personal communication). As of 2019 she was still alive, could recognise her mother's voice, could fixate, vocalised, seemed to experience pain but did not localise the source. The clinicians thought the child could distinguish between environments, could cry and laugh, and play with her mother with a degree of intentionality (rather than displaying purely reactive and automatic behaviours).

There is also another older hydranencephalic case, with reportedly no neo-cortex preserved (as per neuroanatomical investigation following death). In this case, who was in a vegetative state of decorticate rigidity, sensory reinforcement learning was recorded as well as vocalisations and head turning in apparent response to "discomfort"[150]. A modern anencephalic case, who survived much longer than is typical, showed evidence of independent feeding, making cooking noises and smiling spontaneously[151]. The first author recalls "her getting upset and being comforted by her mother" (in personal communication, similar cases have been reported in the news).

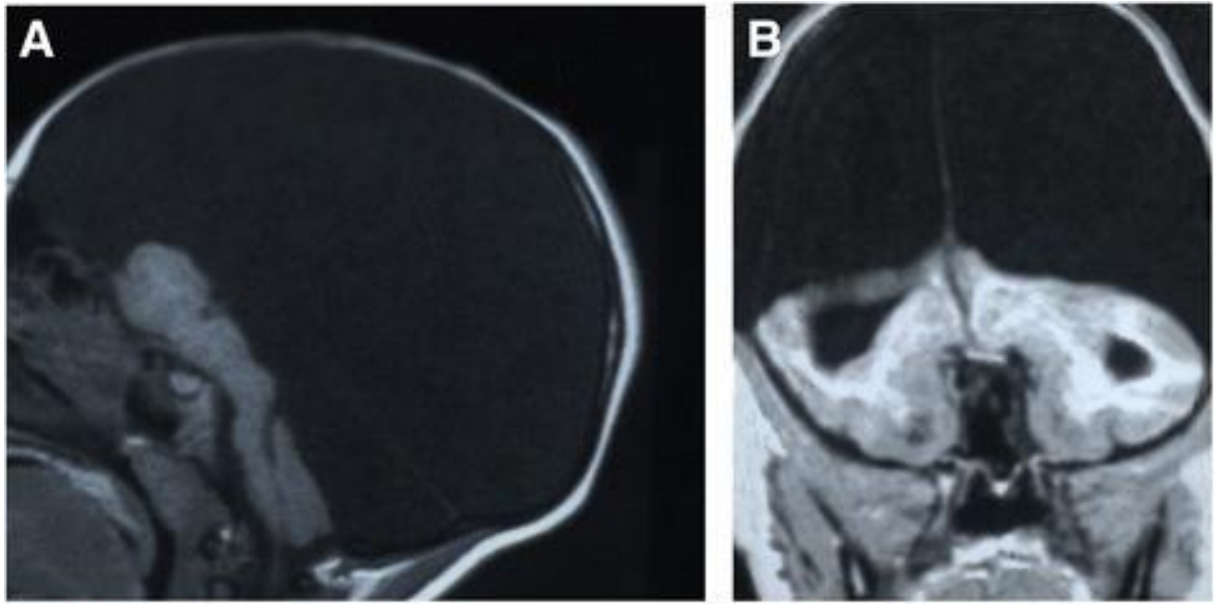

**Figure 5. Brain of hydranencephalic individual with no cortex**. Case of a hydranencephalic child[145] with no cerebral cortex remaining according to clinical neuroradiological assessment. Sagittal (left) and coronal (right) T1 MRI images. Senior author (R.F.) confirmed that the patient did not have any cerebral cortex remaining in personal communication. Clinician who saw the child affirmed that the patient was not in a vegetative state in personal communication. Another affirmed that she "definitely felt emotions". Image reused under CC-BY.

Of note, the behaviours of decorticate individuals may be displays of automatic reactions devoid of experience. It should be repeated, more generally, that that the presence of consciousness is non falsifiable (i.e., presence or absence cannot be conclusively proved, see philosophical zombies thought experiment, locked in syndrome, Descartes' evil genius, etc.,[1,2,13]). However, the evidence regarding hydranencephaly, taken together with the implications arising from neurostimulation (section 3) and neuropsychological evidence (section 4), suggests that, whilst the cortex is important for the full NCCs of a typical human, it may not be necessary to experience in the strictest sense (i.e., sine qua non). Rather, in mammals, the subcortex seems to be the only fundamental brain system required for the emergence of basic subjective experience.

However, the hydranencephalic children reported above are congenital cases. The rather surprising functionality of these human cases may be ascribed to plasticity and a reorganisation of function. The question arises whether the removal of the cortex further along development provokes irreversible unconsciousness (perhaps via diaschisis; or development of essential interdependency between systems). The human decorticate cases suggest that the cortex's anatomical structure is not necessary to experience. However, the subcortex, in absence of a cortex, may have acquired "cortical functions" (such as those enabling the emergence of experience) during such atypical development. Hence, the following section is dedicated to studies which performed an experimental removal of the cortex in animals. Such reports may provide further clues as to what structures may be necessary to consciousness.

### 4.5. Experimental decortication in animals

Similarly to a number of observations in different animals reported by William James over a century ago[152], experimentally decorticated rats are surprisingly comparable to control rats. In rats with no cortex, there are reports (reviewing decades of research[153,154]) of successful mating and associated rituals (variably across individuals), sufficiently preserved maternal behaviour, grooming behaviours, preserved simple association learning (although failing in more complex types of learning) and normal postures. Furthermore, completely decorticate rats showed essentially the same playing behaviour as control rats (with minor differences)[155]. On the other hand, deficits are reported in skilled behaviours (e.g., grooming, although these can improve with learning), and hoarding[154].

Decorticate cats displayed a comparable range of complex behaviours[156] and are indistinguishable to other cats in the first month of life. Play, sexual[157], grooming, emotional and maternal behaviours were observed[156]. The decorticate cats made use of haptic, auditory and visual senses (although the latter was impaired; especially so in adult preparations[158]). Similarly to other observations in cats and dogs[141] they display full, if not heightened, aggressive responses. Some behavioural differences (mostly to do with social behaviour) were observed and these became more prominent as time went by. Partially preserved learning capabilities in these animals are also documented[136,156]. In hypothalamus preparations (removing all rostral matter), there are reports of a full behavioural expression of anger[141,142,158], in contrast to decerebrate cats (i.e., brainstem truncations). This evidence does not prove unequivocally that these experimentally decorticated animals retained subjective experience (an epistemic problem that persists clinically with humans), however, they displayed complex behaviours which, parsimoniously, are likely associated with certain experiences[141].

However, of primary importance is whether surgical decortication performed later in development would induce irreversible unconsciousness (given the differences between the congenital and acquired cases reviewed, and possible reorganisation of function). Reports suggest that behavioural deficits are surprisingly comparable between rats and cats that received complete decortication neonatally and in adulthood[154,159,160]. Neonatal lesioning leads to improved performance in a variety of behaviours (e.g., maternal behaviour) compared to adult decortication. As expected, decortication provokes secondary changes to subcortical structures (e.g., necrosis; diaschisis[161], as evaluated in post-mortem) which differ between adult or neonatally decorticated rats[154,159,160]. Nonetheless, animals that had their cortex removed in adulthood seemed to display behaviours that may suggest the presence of experience (e.g., fear, anger, restlessness, eating, sexual behaviour; certain forms of learning, excessive curiosity; search for human companionship; although expressions of joy are mostly reported to be abolished)[141,154,158,159,162–165]. This evidence suggests abnormal compensatory development cannot wholly explain the behaviours that decorticate animal's display.

The impossibility of probing verbal report in such organisms reduces the certainty that such decorticate animals and humans did experience (anything at all). It must be said, these might effectively be "zombie" rats and cats[1,2], displaying mechanical behaviours without any experience. The removal of the cortex likely has a major effect on the experience of these animals. Nonetheless, if behaviours can be taken to suggest experience, as is the gold standard in the clinic[9], the evidence reviewed above tentatively suggests that the subcortex is sufficient for the emergence of a basic form of experience.

### 4.5.1 Primate decortication

Nonetheless, such animals may not be comparable to primates. Although behavioural evidence of basic forms of experience does exit when these rats, cats and dogs have their cortex removed; primates, which have phylogenically and ontogenically highly developed cortexes may perhaps have a different mechanism underlying the generation of consciousness (as suggested by Penfield and Jaspers[143]). There are in fact very few obtainable studies to my knowledge that look at experimentally decorticated primates. This is partly due to the difficulty of performing such surgery with the primate surviving and recovering sufficiently[166,167], but also the ethical dubiousness of such experiments (which may suggest the possibility of experiences of suffering). In these studies, descriptions tend to focus on motor function and do not give much behavioural detail.

With a surgical procedure involving two sequential hemispherectomies, Kennard describes two decorticated Macaques showing strong startle responses to auditory stimuli. Two other infant-decorticated monkeys showed appropriate vocalisations to pain as well as other startle responses somewhat suggestive of "emotional expression"[168] (p. 296). However, in these cases, the decorticate monkeys did not survive long after surgery. Karpus and Kreidl in 1914[161], report 7 cases that survived more than 24 hours after two sequential hemispherectomies (where reported, animals were approximately 1 year of age). Out of these seven, only one was observed to substantially recover its functions, whilst the others displayed the highest functionality the day after the second hemispherectomy. Importantly, the authors sometimes report finding small remnants of cortex during the autopsy (e.g., gyrus uneinnatus, orbital gyrus, gyrus fornicates). They write of clear transitions between sleep-like and wakeful states. In the former state, animals did not perform spontaneous movements, did not react to stimuli or vocalise, and had their eyes shut. They describe one decorticated individual who used its limbs on the cage bars to right itself into a seated position. They report occasional behavioural responses to auditory, tactile and visual stimuli (eye opening and movements, limb movements), "cries of pain" (but often no facial expression), as well as other vocalisations. In another case the totally decorticated monkey removed a cotton bandage, emitted vocal sounds and cries, pulls itself up with the bars of the cage, and swallowed food and drinks. After the second hemispherectomy, they report "there are no clear signs of change of the state of consciousness". Although their lack of specification of their conception of consciousness is problematic, fewer behavioural signs have been interpreted as indicative of consciousness in recent, high impact, animal-model literature[103,104]. Another macaque (shown in figure 6) showed vigorous movement, alternating periods of sleep and wakefulness in which it positions the body via use of limbs (e.g. seated, righting reflex often used as an indication of consciousness in animal studies[107]). It moved around the whole cage, and emitted vocalisations comparable to a healthy monkey although with a striking lack of facial expressions. It showed behavioural responses to tactile and auditory stimuli although individual could not feed itself independently.

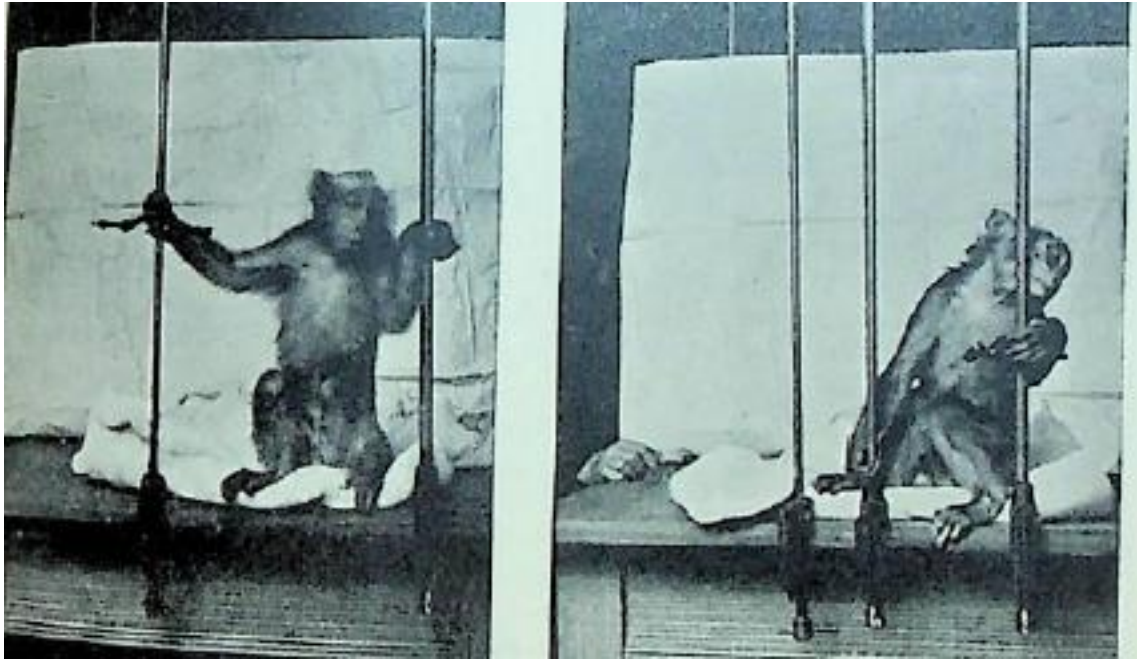

**Figure 6. Picture of decorticated monkey in Karpus and Kreidl (1914).** Right panel would show the monkey in a more "awake" state; left panel in a more "sleepy" state. Note, reportedly, a portion of basal ganglia was also removed, although minimal remnants of cortical gyrus uneinnatus and substatia perforate remained, as assessed in post mortem; Copyright expired.

Another case (surviving 26 days) shows evidence of feeding, startle reaction to acoustic (e.g., call from another monkey) and tactile stimuli, righting response and vocalisations. It walked and hopped and spontaneously scratched itself. It started recovering strength and had lively responses when manipulated until it succumbed to dysentery. They explicitly comment that they cannot draw conclusions as to the decorticated monkey, due to a lack of maximum recovery in these animals.

A decade later, Travis and Woolsey[167] succeeded in fully decorticating (neocortex) two *adult* macaques via serial surgery. Whilst one required help, one of the monkeys was able to right itself independently (traditionally, as a sign of consciousness in animal research[138]). This monkey could also walk on its own (showing a degree of "caution", being blind). Similarly to above, the authors also suggest that the monkeys in question were not permitted to recover to their full potential[167].

The lack of research and relevant detail on the available cases does not permit a strong conclusion. Although there seems to be a difference between phylogenic and ontogenic development on the effect of decortication, there is still a remarkable capability of function retained across decorticated animals. Importantly, however, these behavioural descriptions may be reasonable cause for doubt that the near complete destruction of the cortex in adult primates is unequivocally associated with unconsciousness.

## 5. Discussion of subcortical function

To draw the evidence together, I reviewed how all major structures of the brain display correlations with consciousness in neuroimaging studies. In neurostimulation and neuropsychological studies, changes to cortical and cerebellar regions were related to quite substantial changes in experience. However, the subcortex showed the strongest effects. Taking into account various epistemic problems, a review of animal and human decorticated cases indicates that the subcortex is sufficient for some of the behaviours that are considered correlates of consciousness.

Panksepp[94] suggests that such subcortical structures, which tend to be evolutionarily conserved across species, have the function of imbuing the animal with evaluative affects in regards to basic physiological functions (as well as regulating such functions). These primary "value" feelings provide the animal with crucial information regarding its survival (e.g., hunger vs satiety). Such responses to value-laden stimuli can be described as automatic ("unconditioned response" in behavioural psychology). Such fundamental physiological responses may be accompanied by basic feelings or emotions regulating survival-related behaviours, which can be described as affective, interoceptive and homeostatic[34,94,169]. Such affective information underlies behaviours motivated by sexual lust, feelings of cold, pain, hunger, the need to breath, play, attachment, and many others. Thus, they are essentially a representation of the state of the system to the system itself[34,116]. Importantly, these subcortical basic responses likely mediate reinforcement learning in the environment[97,150,156] and can therefore lead to complex behaviours (e.g., in decorticate cases[156]). Whilst accounts of consciousness that tend to focus on the cortex emphasise higher forms of awareness (e.g., regarding the self, access consciousness; working memory), the most basic forms of qualitative experiences, primary affects, may not be fundamentally reliant on the cortex.

Different authors have thought that such basic feelings were the first experiences to be engendered in the history of our world[31,94], and perhaps the lowest level of consciousness[34,116]. Of note is that these simple affects/feelings and physiological reactions energize behaviour (e.g., hunger causes one to find food) and are of extreme importance to life[32]. We can therefore understand such primary affects in the context of the free energy principle[170]. Such feelings cause a discomfort that have the overall function of reducing free energy ("surprising" states that are not conducive to bodily function over the long term). On this view, subjective experience has an origin that is very much pertaining to survival. Hence, I will put forth the proposition, that whilst the front or the back of the cortex and the cerebellum, are an essential part of the "full" neural correlates of consciousness, the sufficient neural conditions of fundamental experiences likely lie in the subcortex. Here, I propose a distinction which is not conclusive; but may form the basis for hypotheses in future investigations.

***Primary forms of consciousness*** may be as simple as a feeling of pain or basic interoception[19,31,116,136,171]. These subjective feelings, in mammals at least, seem to rely on subcortical structures. ***Higher-order forms of consciousness*** may be complex processes such as working memory associated functions[20,171], self-referential and prospective thought, verbal report and representation, poetic sentiments and wilful behaviour mediated by executive decisions. Although many higher-order forms of consciousness would plausibly depend on the cortex (as well as the cerebellum), I put forth the hypothesis that higher-order forms of consciousness depend on primary forms which arise out of older neural systems and are more fundamental to life. Such a mechanism may fundamentally imply a dynamic interaction between subcortical structures, the body and the environment[5,31,32,136,172]. As Panksepp argues[94], the subcortex "underlied the first subjective experiences in the world". In fact, Damasio and Carvalho propose[116] that the first feelings emerged even before the phylogenic advent of the cerebral cortex.  However, such cortical remapping of basic, subcortically instantiated, interoceptive and affective information would provide higher-order, more finely discriminative functionalities (integrating, for example, bodily alloceptive and proprioceptive coordinates, cognitive expectation and self-appraisal to a simple emotion in relation to a social environment etc.)[116].

Such notions of a primary, minimal phenomenological type of consciousness is found in the work of both consciousness scientists and philosophers[31,32,171]. There are many subtle distinctions and different possible formulations of such "pre-reflective" primary forms of awareness. Of importance to consciousness science, the various empirical and theoretical difficulties seen above[4] are exacerbated by the lack of a clear definition and operationalisation of consciousness; as well as a lack of reporting of relevant behavioural detail. Future work may have to identify the differences in conceptual formulations and try to match them with empirical data. Although it permits interesting thoughts, the evidence reviewed here does not seem to provide conclusive evidence in favour of any of the major theories of consciousness (e.g., brainstem is in an excellent position to integrate information, broadcast it, for re-entrant loops, and re-representation[3,22,24,28,173]). Either way, if the subcortex is truly sufficient for experience to emerge, then investigations may be undertaken to narrow down what specific functions engendered by the subcortex are necessary for experience, thus making way on the hard problem of consciousness[1]. Clinical and ethical implications may arise out of such notions, but this is beyond the scope of this article.

## 6. Intrinsic and extrinsic context as a solution to empirical paradoxes

There is still a fundamental problem to solve arising out of all the evidence reviewed above. The cortex, subcortex and cerebellum may all display correlations with consciousness. When reviewing causal evidence, there seems to be a causal role of the subcortex in consciousness. In particular, there is the case of hydranencephalic children[147], which, with little or no cortex, seem to be able to experience. Yet, I cannot ignore the evidence of rather extreme (reported) changes to subjective experience with the alteration of frontal and posterior midline cortices[26,81]. Similarly, whilst cases of cerebellar agenesis[121] show that the cerebellum may not be necessary for consciousness in the absolute sense, in cases of posterior fossa syndrome or cerebellar cognitive affective syndrome, quite severe changes to experience can be reported[77,124,125]. There is, in fact, a clear difference between cases of hydranencephaly[147] compared to posterior midline stimulation[81]; of cerebellar agenesis[121] versus cerebellar damage[77,124,125]; of frontal lobe agenesis[118] contrasted with frontal infarct in an adult[26]. In one case the abnormality is congenital, in another it is acquired (respectively). This highlights the importance of (both internal and external) context of the individual for experience. The destruction of brain regions, once the individual is fully developed (and systemic functionality is subtly calibrated across many dimensions) may provoke great functional obliteration and a considerable impairment to normal consciousness. The rest of the system will be missing something through which it has always functioned[144]. Conversely, a system (without significant parts that would "typically" be there; i.e., abnormal), would adapt and develop intrinsically, within that context, regardless of extrinsic expectations or evaluation. Thus, although the evidence in this review seems to suggest a hierarchy in the importance in "consciousness imbuing" structures, the evidence reviewed here, when taken together, also shows that these systems cannot understood in isolation.

Perhaps the Sprague effect, as suggested by Merker[137] (who's work is foundational to this article), illustrates the mutual dependence between the cortex and the subcortex in normal function. After occipital cortical damage there is a loss of visual function. However, astonishingly, some visual function is restored by additional damage to the superior colliculi[136,174] in the brainstem. This is interpreted as the absence of cortical input provoking a secondary brainstem effect preventing its normal visual processing function (e.g., via excessive inhibition, diaschisis). Thus, normal brainstem function may be impaired by the absence of an input through which its function was calibrated, and restored to an extent once the need for this input is destroyed (see also Snider et al., 2020).

To put it in the language complex systems, via the destruction of parts of, the systemic functioning of the whole is severely impaired. On the other hand, in congenital cases, the brain body and environment interactions[172] can develop within their own "atypical" context, thus permitting the representation of basic forms of interoceptive and homeostatic self and perhaps the development of more complex types of experience over time (e.g., via presumed operant conditioning in hydranencephalic individuals). The importance of the temporal and physical context in the emergence of consciousness is perhaps evidenced by the fact that hydranencephalic individuals who were given species-typical environmental interactions, displayed behavioural signs that would not conform to a vegetative state presentation (in opposition to those who were not[147]). Beyond the fundamental physiological context of the **body** and its various structures, this furthermore highlights the temporal context, not only in terms of biology (e.g., development, tonic modulation) but also from a phenomenological point of view: in the stream of consciousness, the present must have its own past, and which will lead to a future[34,53,173].

## Acknowledgements

Thank you kindly to Hanna Tolle for substantial help in translating Karplus and Kreidl, 1914. Thank you to Professor Pavone for several meetings regarding the hydranencephalic case reported in Pavone et al., 2014. I would also like to acknowledge Professor Enrico Parano for his comments on the same case. I would also like to acknowledge Dr Holly Dickman, for comments regarding the anencephalic patient reported in Dickman et al., 2016. I would like to kindly thank Professor Antonio Damasio for his comments and support. Thank you to Dr Dian Lyu for various discussion and comments. I would also like to acknowledge Dr Lenni Spindler for various discussion surrounding this topic over the years. My gratitude is also due to David Milton. I would also like to acknowledge the Cambridge University librarians who have patiently endeavoured to find articles relevant to this work. An acknowledgment is due to the Cambridge Trust who funded my doctoral studies from which this review was spawned.

# References

1. Chalmers, D. Facing up to the problem of consciousness. *J. Conscious. Stud.* **2**, 121–152 (1995).

2. Dennet, D. *Consciousness Explained*. (Penguin, 1991).

3. Tononi, G., Boly, M., Massimini, M. & Koch, C. Integrated information theory: From consciousness to its physical substrate. *Nat. Rev. Neurosci.* **17**, 450–461 (2016).

4. Koch, C., Massimini, M., Boly, M. & Tononi, G. Neural correlates of consciousness: Progress and problems. *Nat. Rev. Neurosci.* **17**, 307–321 (2016).

5. Klein, C. & Barron, A. B. Insects have the capacity for subjective experience. *Anim. Sentience* **1**, (2016).

6. Nagel, T. What Is It Like to Be a Bat? *Philos. Rev.* **83**, (1974).

7. Sanders, R. D., Tononi, G., Laureys, S. & Sleigh, J. W. Unresponsiveness ≠ unconsciousness. *Anesthesiology* **116**, 946–959 (2012).

8. Edlow, B. L., Claassen, J., Schiff, N. D. & Greer, D. M. Recovery from disorders of consciousness: mechanisms, prognosis and emerging therapies. *Nat. Rev. Neurol.* **17**, 135–156 (2021).

9. Giacino, J. T., Kalmar, K. & Whyte, J. The JFK Coma Recovery Scale-Revised: Measurement

characteristics and diagnostic utility11No commercial party having a direct financial interest in the results of the research supporting this article has or will confer a benefit upon the authors or upon an. *Arch. Phys. Med. Rehabil.* **85**, 2020–2029 (2004).

10. Laureys, S., Perrin, F. & Brédart, S. Self-consciousness in non-communicative patients. *Conscious. Cogn.* **16**, 722–741 (2007).

11. Russell, I. F. Fourteen fallacies about the isolated forearm technique, and its place in modern anaesthesia. *Anaesthesia* **68**, 677–681 (2013).

12. Fu, V. X. *et al.* Perception of auditory stimuli during general anesthesia and its effects on patient outcomes: a systematic review and meta-analysis. *Can. J. Anesth.* **68**, 1231–1253 (2021).

13. Owen, A. M. & Coleman, M. R. Detecting awareness in the vegetative state. *Ann. N. Y. Acad. Sci.* **1129**, 130–138 (2008).

14. Posner, J. B., Saper, C. B., Schiff, N. D. & Plum, F. *Diagnosis of stupor and coma*. *Plum and Posner's Diagnosis of Stupor and Coma* (2013). doi:10.1093/med/9780195321319.003.0010

15. Cortese, M. D. *et al.* Coma recovery scale-r: Variability in the disorder of consciousness. *BMC Neurol.* **15**, 1–7 (2015).

16. Wannez, S., Heine, L., Thonnard, M., Gosseries, O. & Laureys, S. The repetition of behavioral assessments in diagnosis of disorders of consciousness. *Ann. Neurol.* **81**, 883–889 (2017).

17. Schnakers, C. *et al.* Diagnostic accuracy of the vegetative and minimally conscious state: Clinical consensus versus standardized neurobehavioral assessment. *BMC Neurol.* **9**, 1–5 (2009).

18. Odor, P. M. *et al.* Incidence of accidental awareness during general anaesthesia in obstetrics: a multicentre, prospective cohort study. *Anaesthesia* **76**, 759–776 (2021).

19. Panksepp, J., Fuchs, T., Garcia, V. A. & Lesiak, A. Does any aspect of mind survive brain damage that typically leads to a persistent vegetative state? Ethical considerations. *Philos. Ethics, Humanit. Med.* **2**, 1–11 (2007).

20. Crick, F. & Koch, C. Towards a neurobiological theory of consciousness. *Semin. neurosceinces* **2**, 263–275 (1990).

21. Boly, M. *et al.* Are the neural correlates of consciousness in the front or in the back of the cerebral cortex? Clinical and neuroimaging evidence. *J. Neurosci.* **37**, 9603–9613 (2017).

22. Lau, H. & Rosenthal, D. Empirical support for higher-order theories of conscious awareness. *Trends Cogn. Sci.* **15**, 365–373 (2011).

23. Brown, R., Lau, H. & LeDoux, J. E. Understanding the Higher-Order Approach to Consciousness. *Trends Cogn. Sci.* **23**, 754–768 (2019).

24. Seth, A. K. & Bayne, T. Theories of consciousness. *Nat. Rev. Neurosci.* **23**, 439–452 (2022).

25. Pennartz, C. M. A., Farisco, M. & Evers, K. Indicators and criteria of consciousness in animals and intelligent machines: An inside-out approach. *Front. Syst. Neurosci.* **13**, 1–23 (2019).

26. Odegaard, B., Knight, R. T. & Lau, H. Should a few null findings falsify prefrontal theories of conscious perception? *J. Neurosci.* **37**, 9593–9602 (2017).

27. Melloni, L., Mudrik, L., Pitts, M. & Koch, C. Making the hard problem of consciousness easier. *Science (80-. ).* **372**, 911–912 (2021).

28. Baars, B. J. Global workspace theory of consciousness: Toward a cognitive neuroscience of human experience. *Prog. Brain Res.* **150**, 45–53 (2005).

29. Alkire, M. T., Hudetz, A. G. & Tononi, G. Consciousness and Anesthesia NIH Public Access. *Science (80-. ).* **322**, 876–880 (2008).

30. Golkowski, D. *et al.* Changes in Whole Brain Dynamics and Connectivity Patterns during Sevoflurane- A nd Propofol-induced Unconsciousness Identified by Functional Magnetic Resonance Imaging. *Anesthesiology* **130**, 898–911 (2019).

31. Damasio, A. R. Investigating the biology of consciousness. *Philos. Trans. R. Soc. B Biol. Sci.* **353**, 1879–1882 (1998).

32. Parvizi, J. & Damasio, A. Consciousness and the brainstem. *Cognition* **79**, 135–160 (2001).

33. Solms, M. The conscious Id. *Neuropsychoanalysis* **15**, 5–19 (2013).

34. Damasio, A. & Damasio, H. Homeostatic feelings and the biology of consciousness. *Brain* **145**, 2231–2235 (2022).

35. Demertzi, A. *et al.* Human consciousness is supported by dynamic complex patterns of brain signal coordination. *Sci. Adv.* **5**, (2019).

36. Achard, S. *et al.* Hubs of brain functional networks are radically reorganized in comatose patients. *Proc. Natl. Acad. Sci.* **109**, 20608–20613 (2012).

37. Tagliazucchi, E. *et al.* Breakdown of long-range temporal dependence in default mode and

attention networks during deep sleep. *Proc. Natl. Acad. Sci. U. S. A.* **110**, 15419–15424 (2013).

38. Dupont, M. *et al.* Resting-state Dynamics as a Cortical Signature of Anesthesia in Monkeys. *Anesthesiology* **129**, 942–958 (2018).

39. Barttfeld, P. *et al.* Signature of consciousness in the dynamics of resting-state brain activity. *Proc. Natl. Acad. Sci. U. S. A.* **112**, 887–892 (2015).

40. Naci, L., Sinai, L. & Owen, A. M. Detecting and interpreting conscious experiences in behaviorally non-responsive patients. *Neuroimage* **145**, 304–313 (2017).

41. Deng, F., Taylor, N., Owen, A. M., Cusack, R. & Naci, L. The Brain-Bases of responsiveness variability under moderate anaesthesia. **353**, (2020).

42. Cavanna, F., Vilas, M. G., Palmucci, M. & Tagliazucchi, E. Dynamic functional connectivity and brain metastability during altered states of consciousness. *NeuroImage* **180**, 383–395 (2018).

43. Bastos, A. M. *et al.* Neural effects of propofol-induced unconsciousness and its reversal using thalamic stimulation. *Elife* **10**, 1–28 (2021).

44. Huang, Z., Zhang, J., Wu, J., Mashour, G. A. & Hudetz, A. G. Temporal circuit of macroscale dynamic brain activity supports human consciousness. *Sci. Adv.* **6**, 1–15 (2020).

45. Varley, T., Denny, V., Sporns, O. & Patania, A. Topological Analysis of Differential Effects of Ketamine and Propofol Anesthesia on Brain Dynamics. 1–25 (2020). doi:10.1101/2020.04.04.025437

46. Hutchison, R. M., Hutchison, M., Manning, K. Y., Menon, R. S. & Everling, S. Isoflurane induces dose-dependent alterations in the cortical connectivity profiles and dynamic properties of the brain's functional architecture. *Hum. Brain Mapp.* **35**, 5754–5775 (2014).

47. Huang, Z., Liu, X., Mashour, G. A. & Hudetz, A. G. Timescales of intrinsic BOLD signal dynamics and functional connectivity in pharmacologic and neuropathologic states of unconsciousness. *J. Neurosci.* **38**, 2304–2317 (2018).

48. Guldenmund, P., Vanhaudenhuyse, A., Boly, M., Laureys, S. & Soddu, A. A default mode of brain function in altered states of consciousness. 107–121 (2012).

49. Vatansever, D., Menon, D. K. & Stamatakis, E. A. Default mode contributions to automated information processing. *Proc. Natl. Acad. Sci. U. S. A.* **114**, 12821–12826 (2017).

50. Di Perri, C. *et al.* Neural correlates of consciousness in patients who have emerged from a

minimally conscious state: A cross-sectional multimodal imaging study. *Lancet Neurol.* **15**, 830–842 (2016).

51. Mashour, G. A. *et al.* Recovery of consciousness and cognition after general anesthesia in humans. *Elife* **10**, 1–21 (2021).

52. Luppi, A. I. *et al.* Consciousness-specific dynamic interactions of brain integration and functional diversity. *Nat. Commun.* **10**, (2019).

53. Coppola, P. *et al.* The complexity of the stream of consciousness. *Commun. Biol.* **5**, 1–15 (2022).

54. Demertzi, A. *et al.* Intrinsic functional connectivity differentiates minimally conscious from unresponsive patients. *Brain* **138**, 2619–2631 (2015).

55. Vanhaudenhuyse, A. *et al.* Default network connectivity reflects the level of consciousness in non-communicative brain-damaged patients. *Brain* **133**, 161–171 (2010).

56. Guldenmund, P. *et al.* Thalamus, Brainstem and Salience Network Connectivity Changes During Propofol-Induced Sedation and Unconsciousness. *Brain Connect.* **3**, 273–285 (2013).

57. Boveroux, P. Breakdown of within- and between-network Resting State during Propofol-induced Loss of Consciousness. *Anesthesiology* **113**, 1038–1053 (2010).

58. Monti, M. M. *et al.* Dynamic Change of Global and Local Information Processing in Propofol-Induced Loss and Recovery of Consciousness. *PLoS Comput. Biol.* **9**, (2013).

59. Amico, E. *et al.* Posterior cingulate cortex-related co-activation patterns: A resting state fMRI study in propofol-induced loss of consciousness. *PLoS One* **9**, (2014).

60. Coppola, P. *et al.* Network dynamics scale with levels of awareness. *Neuroimage* **254**, 119128 (2022).

61. Gili, T. *et al.* The Thalamus and Brainstem Act As Key Hubs in Alterations of Human Brain Network Connectivity Induced by Mild Propofol Sedation. *J. Neurosci.* **33**, 4024–4031 (2013).

62. Tsurugizawa, T. & Yoshimaru, D. Impact of anesthesia on static and dynamic functional connectivity in mice. *Neuroimage* **241**, 118413 (2021).

63. Crone, J. S. *et al.* Altered network properties of the fronto-parietal network and the thalamus in impaired consciousness. *NeuroImage Clin.* **4**, 240–248 (2014).

64. Guldenmund, P. *et al.* Brain functional connectivity differentiates dexmedetomidine from

propofol and natural sleep. *Br. J. Anaesth.* **119**, 674–684 (2017).

65. Stamatakis, E. A., Adapa, R. M., Absalom, A. R. & Menon, D. K. Changes in resting neural connectivity during propofol sedation. *PLoS One* **5**, (2010).

66. Arthuis, M. *et al.* Impaired consciousness during temporal lobe seizures is related to increased long-distance corticalsubcortical synchronization. *Brain* **132**, 2091–2101 (2009).

67. Mhuircheartaigh, R. N. *et al.* Cortical and subcortical connectivity changes during decreasing levels of consciousness in humans: A functional magnetic resonance imaging study using propofol. *J. Neurosci.* **30**, 9095–9102 (2010).

68. Spindler, L. R. B. *et al.* Dopaminergic brainstem disconnection is common to pharmacological and pathological consciousness perturbation. *Proc. Natl. Acad. Sci. U. S. A.* **118**, 1–11 (2021).

69. Fernández-Espejo, D. *et al.* Diffusion weighted imaging distinguishes the vegetative state from the minimally conscious state. *Neuroimage* **54**, 103–112 (2011).

70. Lutkenhoff, E. S., Johnson, M. A., Casarotto, S., Massimini, M. & Monti, M. M. Subcortical atrophy correlates with the perturbational complexity index in patients with disorders of consciousness. *Brain Stimul.* **13**, 1426–1435 (2020).

71. Lutkenhoff, E. S. *et al. The subcortical basis of outcome and cognitive impairment in TBI: A longitudinal cohort study*. *Neurology* **95**, (2020).

72. Smallwood, J. *et al.* The default mode network in cognition: a topographical perspective. *Nat. Rev. Neurosci.* **22**, 503–513 (2021).

73. Block, N. What Is Wrong with the No-Report Paradigm and How to Fix It. *Trends Cogn. Sci.* **23**, 1003–1013 (2019).

74. Keren Happuch, E., Chen, S. H. A., Ho, M. H. R. & Desmond, J. E. A meta-analysis of cerebellar contributions to higher cognition from PET and fMRI studies. *Hum. Brain Mapp.* **35**, 593–615 (2014).

75. Stoodley, C. J. & Schmahmann, J. D. Functional topography in the human cerebellum: A meta-analysis of neuroimaging studies. *Neuroimage* **44**, 489–501 (2009).

76. Gatti, D., Rinaldi, L., Cristea, I. & Vecchi, T. Probing cerebellar involvement in cognition through a meta-analysis of TMS evidence. *Sci. Rep.* **11**, 1–17 (2021).

77. Adamaszek, M. *et al.* Consensus Paper: Cerebellum and Emotion. *Cerebellum* **16**, 552–576

(2017).

78. Baumann, O. *et al.* Consensus Paper: The Role of the Cerebellum in Perceptual Processes. *Cerebellum* **14**, 197–220 (2015).

79. Mariën, P. *et al.* Consensus paper: Language and the cerebellum: An ongoing enigma. *Cerebellum* **13**, 386–410 (2014).

80. Selimbeyoglu, A. & Parvizi, J. Electrical stimulation of the human brain: Perceptual and behavioral phenomena reported in the old and new literature. *Front. Hum. Neurosci.* **4**, 1–11 (2010).

81. Parvizi, J. *et al.* Altered sense of self during seizures in the posteromedial cortex. *Proc. Natl. Acad. Sci. U. S. A.* **118**, 1–9 (2021).

82. Lyu, D. *et al.* Causal evidence for the processing of bodily self in the anterior precuneus. *Neuron* 1–11 (2023). doi:10.1016/j.neuron.2023.05.013

83. Herbet, G. *et al.* Disrupting posterior cingulate connectivity disconnects consciousness from the external environment. *Neuropsychologia* **56**, 239–244 (2014).

84. Rounis, E., Maniscalco, B., Rothwell, J. C., Passingham, R. E. & Lau, H. Theta-burst transcranial magnetic stimulation to the prefrontal cortex impairs metacognitive visual awareness. *Cogn. Neurosci.* **1**, 165–175 (2010).

85. Xu, Z. *et al.* Behavioral Effects in Disorders of Consciousness Following Transcranial Direct Current Stimulation : A Systematic Review and Individual Patient Data Meta-analysis of Randomized Clinical Trials. *MedRxiv* (2021).

86. Xia, X. *et al.* Current Status of Neuromodulatory Therapies for Disorders of Consciousness. *Neurosci. Bull.* **34**, 615–625 (2018).

87. Bourdillon, P., Hermann, B., Sitt, J. D. & Naccache, L. Electromagnetic Brain Stimulation in Patients With Disorders of Consciousness. *Front. Neurosci.* **13**, 1–14 (2019).

88. Raccah, O., Block, N. & Fox, K. C. R. Does the prefrontal cortex play an essential role in consciousness? insights from intracranial electrical stimulation of the human brain. *J. Neurosci.* **41**, 2076–2087 (2021).

89. Fox, K. C. R. *et al.* Intrinsic network architecture predicts the effects elicited by intracranial electrical stimulation of the human brain. *Nat. Hum. Behav.* **4**, 1039–1052 (2020).

90. Ferrari, C. *et al.* TMS over the Cerebellum Interferes with Short-term Memory of Visual Sequences. *Sci. Rep.* **8**, 1–8 (2018).

91. Cattaneo, Z. *et al.* Cerebellar vermis plays a causal role in visual motion discrimination. *Cortex* **58**, 272–280 (2014).

92. Ferrari, C., Ciricugno, A., Urgesi, C. & Cattaneo, Z. Cerebellar contribution to emotional body language perception: A TMS study. *Soc. Cogn. Affect. Neurosci.* **2019**, 81–90 (2021).

93. Naro, A. *et al.* Cortical connectivity modulation induced by cerebellar oscillatory transcranial direct current stimulation in patients with chronic disorders of consciousness: A marker of covert cognition? *Clin. Neurophysiol.* **127**, 1845–1854 (2016).

94. Panksepp, J. The basic emotional circuits of mammalian brains: Do animals have affective lives? *Neurosci. Biobehav. Rev.* **35**, 1791–1804 (2011).

95. Bejjani, B. *et al.* Transient acute depression induced by high-frequency deep-brain stimulation. *N. Engl. J. Med.* 1476–1480 (1999).

96. Kuhn, J. *et al.* Deep brain stimulation of the nucleus accumbens and its usefulness in severe opioid addiction. *Mol. Psychiatry* **19**, 145–146 (2014).

97. Linden, D. *The Compass of Pleasure: How Our Brains Make Fatty Foods, Orgasm, Exercise, Marijuana, Generosity, Vodka, Learning, and Gambling Feel So Good.* (2011).

98. Parvizi, J. *et al.* Complex negative emotions induced by electrical stimulation of the human hypothalamus. *Brain Stimul.* **15**, 615–623 (2022).

99. Inman, C. S. *et al.* Human amygdala stimulation effects on emotion physiology and emotional experience. *Neuropsychologia* **145**, (2020).

100. Inman, C. S. *et al.* Direct electrical stimulation of the amygdala enhances declarative memory in humans. *Proc. Natl. Acad. Sci. U. S. A.* **115**, 98–103 (2018).

101. Schiff, N. D. *et al.* Behavioural improvements with thalamic stimulation after severe traumatic brain injury. *Nature* **448**, 600–603 (2007).

102. Monti, M. M., Schnakers, C., Korb, A. S., Bystritsky, A. & Vespa, P. M. Non-Invasive Ultrasonic Thalamic Stimulation in Disorders of Consciousness after Severe Brain Injury: A First-in-Man Report. *Brain Stimul.* **9**, 940–941 (2016).

103. Redinbaugh, M. J. *et al.* Thalamus modulates consciousness via layer-secific control of the

cortex. *Neuron* **106**, 66–75 (2020).

104. Tasserie, J. *et al.* Deep brain stimulation of the thalamus restores signatures of consciousness in a nonhuman primate model. *Sci. Adv.* **8**, 1–17 (2022).

105. Moruzzi, G. & Magoun, H. W. Brain stem reticular formation and activation of the EEG. *Electroencephalogr. Clin. Neurophysiol.* (1949).

106. Xu, Q. *et al.* Medial Parabrachial Nucleus Is Essential in Controlling Wakefulness in Rats. *Front. Neurosci.* **15**, 1–16 (2021).

107. Bian, T. *et al.* Noninvasive Ultrasound Stimulation of Ventral Tegmental Area Induces Reanimation from General Anaesthesia in Mice. *Research* **2021**, 1–13 (2021).

108. Koubeissi, M. Z., Bartolomei, F., Beltagy, A. & Picard, F. Electrical stimulation of a small brain area reversibly disrupts consciousness. *Epilepsy Behav.* **37**, 32–35 (2014).

109. Gabor, A. J. & Peele, T. L. Alterations of behavior following stimulation of the claustrum of the cat. *Electroencephalogr. Clin. Neurophysiol.* **17**, 513–519 (1964).

110. Damasio, A., Damasio, H. & Tranel, D. Persistence of feelings and sentience after bilateral damage of the insula. *Cereb. Cortex* **23**, 833–846 (2013).

111. Kumral, E., Bayam, F. E. & Özdemir, H. N. Cognitive and Behavioral Disorders in Patients with Precuneal Infarcts. *Eur. Neurol.* **84**, 157–167 (2021).

112. Berryhill, M. E., Phuong, L., Picasso, L., Cabeza, R. & Olson, I. R. Parietal lobe and episodic memory: Bilateral damage causes impaired free recall of autobiographical memory. *J. Neurosci.* **27**, 14415–14423 (2007).

113. Del Cul, A., Dehaene, S., Reyes, P., Bravo, E. & Slachevsky, A. Causal role of prefrontal cortex in the threshold for access to consciousness. *Brain* **132**, 2531–2540 (2009).

114. Brickner, R. M. Brain of patient A. after bilateral frontal lobectomy; stratus of frontal-lobe problem. *Arch. Neurol. Psychiatry* **68**, 15–21 (1952).

115. The Multi-Society Task Force. Medical Aspects Of The Persistent Vegetative State. **326**, (1994).

116. Damasio, A. & Carvalho, G. B. The nature of feelings: Evolutionary and neurobiological origins. *Nat. Rev. Neurosci.* **14**, 143–152 (2013).

117. Wilson, B. A. & Wearing, D. Prisoner of consciousness: A state of just awakening following

herpes simplex encephalitis. in *Broken memories: Case studies in memory impairment.* (eds. Campbell, R. & Conway, M. A.) 14–30 (Blackwell Publishing, 1995).

118. Ibáñez, A. *et al.* Early bilateral and massive compromise of the frontal lobes. *NeuroImage Clin.* **18**, 543–552 (2018).

119. Young, B., Ropper, A. & Bolton, C. *Coma and Impaired consciousness: a clinical perspective*. (The Mcgraw-Hill Companies, 1998).

120. Lemon, R. N. & Edgley, S. A. Life without a cerebellum. *Brain* **133**, 649–654 (2010).

121. Yu, F., Jiang, Q. J., Sun, X. Y. & Zhang, R. W. A new case of complete primary cerebellar agenesis: Clinical and imaging findings in a living patient. *Brain* **138**, e353 (2015).

122. Adamaszek, M. *et al.* Neural correlates of impaired emotional face recognition in cerebellar lesions. *Brain Res.* **1613**, 1–12 (2015).

123. Van Overwalle, F., D'aes, T. & Mariën, P. Social cognition and the cerebellum: A meta-analytic connectivity analysis. *Hum. Brain Mapp.* **36**, 5137–5154 (2015).

124. Mariën, P. *et al.* Posterior fossa syndrome in adults: A new case and comprehensive survey of the literature. *Cortex* **49**, 284–300 (2013).

125. Hoche, F., Guell, X., Vangel, M. G., Sherman, J. C. & Schmahmann, J. D. The cerebellar cognitive affective/Schmahmann syndrome scale. *Brain* **141**, 248–270 (2018).

126. Buckner, R. L. The cerebellum and cognitive function: 25 years of insight from anatomy and neuroimaging. *Neuron* **80**, 807–815 (2013).

127. Panksepp, J. & Watt, D. What is basic about basic emotions? Lasting lessons from affective neuroscience. *Emot. Rev.* **3**, 387–396 (2011).

128. Parvizi, J. & Damasio, A. R. Neuroanatomical correlates of brainstem coma. *Brain* **126**, 1524–1536 (2003).

129. Cairns, H. disturbances of consciousness with lesions op the brain-stem and diencephalon. *Brain* **75**, (1952).

130. Rohaut, B. *et al.* Deep structural brain lesions associated with consciousness impairment early after hemorrhagic stroke. *Sci. Rep.* **9**, 1–9 (2019).

131. Fischer, D. B. *et al.* A human brain network derived from coma-causing brainstem lesions. *Neurology* **87**, 2427–2434 (2016).

132. Snider, S. B. *et al.* Cortical lesions causing loss of consciousness are anticorrelated with the dorsal brainstem. *Hum. Brain Mapp.* **41**, 1520–1531 (2020).

133. Zimmerman, K. A. *et al.* The biomechanical signature of loss of consciousness: computational modelling of elite athlete head injuries. *Brain* **146**, 3063–3078 (2022).

134. Hindman, J. *et al.* Thalamic Strokes That Severely Impair Arousal Extend Into the Brainstem. *Ann Nuerol.* **84**, 926–930 (2018).

135. Magoun, H. W. Coma following midbrain lesions in the monkey. *Anat. Rec.* (1948).

136. Merker, B. Consciousness without a cerebral cortex: A challenge for neuroscience and medicine. *Behav. Brain Sci.* **30**, 63–81 (2007).

137. Lindsley, D. B., Bowden, J. W. & Magoun, H. W. Effect upon the EEG of acute injury to the brain stem activating system. *Electroencephalogr. Clin. Neurophysiol.* **1**, 475–486 (1949).

138. Fuller, P., Sherman, D., Pedersen, N. P., Saper, C. B. & Lu, J. Reassessment of the structural basis of the ascending arousal system. *J. Comp. Neurol.* **519**, 933–956 (2011).

139. Ranson, S. W. Somnolence Caused By Hypothalamic Lesions. *Arch. Neurol. Psychiatry* (1939).

140. Gerashchenko, D. *et al.* Hypocretin-2-saporin lesions of the lateral hypothalamus produce narcoleptic-like sleep behavior in the rat. *J. Neurosci.* **21**, 7273–7283 (2001).

141. Bard, P. On emotional expression after decortication with some remarks on certain theoretical views: Part I. *Psychol. Rev.* **41**, 309–329 (1934).

142. Bard, P. & Macht, M. B. *Neurological Basis of Behaviour*. (Ciba Foundation, 1958).

143. Penfield, W. & Jasper, H. *Epilepsy and the Functional Anatomy of the Human Brain.* (Little Brown, 1954).

144. Kennard, A. ELECTROENCEPHALOGRAM OF DECORTICATE MONKEYS. *J. Neurophysiol.* (1943).

145. Pavone, P. *et al.* Hydranencephaly: Cerebral spinal fluid instead of cerebral mantles. *Ital. J. Pediatr.* **40**, 1–8 (2014).

146. Takada, K., Shiota, M., Ando, M., Kimura, M. & Inoue, K. Porencephaly and hydranencephaly: A neuropathological study of four autopsy cases. *Brain Dev.* **11**, 51–56 (1989).

147. Shewmon, D. A., Holmes, G. L. & Byrne, P. A. Consciousness in congenitally decorticate children: Developmental vegetative state as self-fulfilling prophecy. *Dev. Med. Child Neurol.* **41**, 364–374 (1999).

148. Aleman, B. & Merker, B. Consciousness without cortex: A hydranencephaly family survey. *Acta Paediatr. Int. J. Paediatr.* **103**, 1057–1065 (2014).

149. Counter, S. A. Preservation of brainstem neurophysiological function in hydranencephaly. *J. Neurol. Sci.* **263**, 198–207 (2007).

150. Deiker, T. & Bruno, R. D. Sensory reinforcement of eyeblink rate in a decorticate human. *Am. J. Ment. Defic.* **80**, 665–667 (1976).

151. Dickman, H., Fletke, K. & Redfern, R. E. Prolonged unassisted survival in an infant with anencephaly. *BMJ Case Rep.* **2016**, 1–3 (2016).

152. James, W. The Principles of Psychology. 4004 (1890).

153. Whishaw, I. Q., Dyck, R. & Kolb, B. Sparing of two types of hippocampal rhythmical slow activity (RSA, theta) in adult rats decorticated neonatally. *Brain Res. Bull.* **26**, 425–427 (1991).

154. Whishaw, I. Q. The Decorticate Rat. in *The Cerebral Cortex of the Rat* (eds. Kolb, B. & Tees, R.) 239–268 (MIT Press, 1990).

155. Panksepp, J., Normansell, L., Cox, J. F. & Siviy, S. M. Effects of neonatal decortication on the social play of juvenile rats. *Physiol. Behav.* **56**, 429–443 (1994).

156. Bjursten, L. M., Norrsell, K. & Norrsell, U. Behavioural repertory of cats without cerebral cortex from infancy. *Exp. Brain Res.* **25**, 115–130 (1976).

157. Sue Carter, C., Witt, D. M., Kolb, B. & Whishaw, I. Q. Neonatal decortication and adult female sexual behavior. *Physiol. Behav.* **29**, 763–766 (1982).

158. Bard, P. On emotional expression after decortication with some remarks on certain theoretical views: Part II. *Psychol. Rev.* **41**, 424–449 (1934).

159. Kolb, B. & Whishaw, I. Q. Decortication of rats in infancy or adulthood produced comparable functional losses on learned and species-typical behaviors. *J. Comp. Physiol. Psychol.* **95**, 468–483 (1981).

160. Villablance, J. R., Burgess, J. W. & Olmstead, C. E. Recovery of function after neonatal or adult hemispherectomy in cats: I. Time course, movement, posture and sensorimotor tests. *Behav. Brain Res.* **19**, 205–226 (1986).

161. Karplus, J. P. & Kreidl, A. Uber Totalexstirpationen einer un beider Brobhirnhemispharen an Affern (Macacus rhesus). *Arch. Anat. und Physiol.* (1914).

162. Zeliony, G. Effets de l'ablation des hemispheres cerebraux. *Reveu Med.* (1929).

163. Whishaw, I. Q. & Kolb, B. Can male decorticate rats copulate? *Behav. Neurosci.* **97**, 270–279 (1983).

164. Whishaw, I. Q. & Kolb, B. Decortication abolishes place but not cue learning in rats. *Behav. Brain Res.* **11**, 123–134 (1984).

165. Bromiley, R. B. Conditioned responses in a dog after removal of the neocortex. **Journal of**, 102–110 (1947).

166. Woolsey, C. N. A monkey which survived bilateral decortica? tion for 161 days. *Proc. Amer. Soc. exp. Biol* (1943).

167. Travis, A. M. & Woolsey, C. N. Motor performance of monkeys after bilateral partial and total cerebral decortications. *Amer. J. Phys. Med* (1956).

168. Kennard, A, M. Reactions Of Monkeys of various ages to partial and complete decortications. *J. Neuropathol. Exp. Neurol.* **3**, 289–310 (1944).

169. A.D. Craig. How do you feel? Interoception: the sense of the physiological condition of the body. *Nat. Rev. Neurosci.* **3**, 655–666 (2002).

170. Friston, K. The free-energy principle: a unified brain theory? *Nat. Rev. Neurosci.* **11**, 127–138 (2010).

171. Edelman, G. M., Gally, J. A. & Baars, B. J. Biology of consciousness. *Front. Psychol.* **2**, 1–7 (2011).

172. Varela, F., Lachaux, J. P., Rodriguez, E. & Martinerie, J. The brainweb: Phase synchronization and large-scale integration. *Nat. Rev. Neurosci.* **2**, 229–239 (2001).

173. Friston, K. Am i self-conscious? (or does self-organization entail self-consciousness?). *Front. Psychol.* **9**, 1–10 (2018).

174. Wallace, S. F., Rosenquist, A. C. & Sprague, J. M. Recovery from cortical blindness mediated by destruction of nontectotectal fibers in the commissure of the superior colliculus in the cat. *J. Comp. Neurol.* **284**, 429–450 (1989).